# Modeling and interpretation of single-cell proteogenomic data

Andrew Leduc[1], Hannah Harens[1], and Nikolai Slavov[1,2,*]

1 Departments of Bioengineering, Biology, Chemistry and Chemical Biology, Single Cell Proteomics Center, and Barnett Institute, Northeastern University, Boston, MA 02115, USA

2 Parallel Squared Technology Institute, Watertown, MA 02472, USA

*Correspondence: nslavov@northeastern.edu

## Abstract

Biological functions stem from coordinated interactions among proteins, nucleic acids and small molecules. Mass spectrometry technologies for reliable, high throughput single-cell proteomics will add a new modality to genomics and enable data-driven modeling of the molecular mechanisms coordinating proteins and nucleic acids at single-cell resolution. This promising potential requires estimating the reliability of measurements and computational analysis so that models can distinguish biological regulation from technical artifacts. We highlight different measurement modes that can support single-cell proteogenomic analysis and how to estimate their reliability. We then discuss approaches for developing both abstract and mechanistic models that aim to biologically interpret the measured differences across modalities, including specific applications to directed stem cell differentiation and to inferring protein interactions in cancer cells from the buffing of DNA copy-number variations. Single-cell proteogenomic data will support mechanistic models of direct molecular interactions that will provide generalizable and predictive representations of biological systems.

## Introduction

Single-cell transcriptomic and genomic technologies have advanced our understanding of cellular diversity[1–4] and transcriptional regulation[5]. While these technologies excel in identifying cell subpopulations, they are insufficient to model regulatory events that include protein interactions, such as those between DNA and transcription factors or between RNA and RNA binding proteins[6,7]. Modeling such mechanisms requires *proteogenomic* analysis, which we define as joint modeling of nucleic acids, proteins, their interactions and modifications. Proteogenomic approaches have proven useful with bulk data[8], and we project that single-cell proteogenomic modeling will extend their utility by (i) enabling cell-type resolved analysis and (ii) providing more data points (across single cells) and hence more statistical power for data-driven modeling[9] of regulatory mechanisms.

Single-cell proteogenomics will help connect single-cell genomics with the numerous post-transcriptional mechanisms – such as dynamically regulated protein synthesis, degradation, translocation, and post-translational modifications – that shape cellular phenotypes. Understanding these mechanisms requires direct and reliable measurements of proteins and nucleic acids. Such data may allow for interpreting the relative contributions and dynamics between transcriptional and post-transcriptional regulation, leading to a more complete understanding of gene regulation and its impact on functional phenotypes.

Proteogenomic modeling requires accounting for measurement reliability; this is essential for distinguishing biological regulation from measurement noise. Yet, estimating the reliability of single-cell measurements is challenging, especially when using affinity reagents with unknown specificity for their cognate epitopes in the context of the analyzed samples. This challenge applies to commonly used single-cell proteogenomic methods as they rely on antibody-based protein measurements[10–13]. The use of multiple affinity reagents per protein substantially increases specificity of detection but can also limit sensitivity[14]. Uncertainty from potential off-target binding fundamentally limits the reliability of quantitative modeling.

This limitation can be addressed by single-cell tandem mass spectrometry (MS) because it allows for estimating the quantification reliability for a large number of proteins based on the consistency of multiple measurements per peptide per single cell and multiple peptides per protein[15,16]. While the potential of MS to estimate the reliability of protein measurements has been used to assess the extent of post-transcriptional regulation across healthy and diseased human tissues[17–19], it has not yet been applied to single-cell analysis.

Here, we discuss the potential and promises of using single-cell proteogenomics for modeling molecular mechanisms in single cells. First we define two different modes of data collection that support proteogenomic modeling. We then outline different types of measurement error and discuss estimating reliability for single-cell transcriptomic and proteomic data. We then discuss potential single-cell proteogenomic analysis, highlighting the spectrum of abstraction at which molecular mechanisms can be inferred and providing specific suggestions for single-cell proteogenomic analysis that can provide new biological perspectives.

# Types of proteogenomic analysis

Measurements of proteins and nucleic acids can be paired to support proteogenomic modeling. This can be done by measuring proteins and mRNAs in different single cells from the same population and subsequently integrating measurements (Fig 1a; demonstrated in ref.[20]), or by multimodal measurements of multiple modalities in the same cell, including chromatin accessibility, proteins, and mRNAs (Fig. 1b; demonstrated in ref.[21–24]). Integrated analysis has the advantage of being amenable to a large variety of approaches for sample preparation and analysis.

## Integrated datasets

If protein and RNA levels are measured in different cells, the cell clusters can be aligned across datasets via common principal component analysis or by dedicated algorithms[25–28]. Since these algorithms are not optimized for MS protein data, new methods that model the error distributions and missingness patterns of MS data are likely to substantially improve integration. Once cell type clusters or gradients of cell states are aligned and clearly defined, statistical models can quantify deviations between protein and mRNA levels that correspond to post-transcriptional regulation. Such alignment is challenging, and the challenge increases with the desired resolution of alignment[29]. Errors in the alignment may contribute to deviations between RNA and protein levels and must be accounted for in making biological inferences.

## Multimodal datasets

Current multimodal methods for obtaining single-cell MS proteomic and RNA-seq data have limited throughput, but we anticipate that high-throughput methods will be developed, e.g., by splitting cell lysates from single cells or by advanced methods that allow separating proteins and RNAs from a single cell. Despite their differences, both integrated and multimodal approaches may allow for developing rigorous models of transcriptional and post-transcriptional regulation in complex samples. Multimodal measurements with MS may naturally extend to other crucial biomolecules, such as metabolites and lipids [30]. In the case of small molecules, integrated analysis presents a significant challenge as the mapping between data types is more ambiguous compared to the relationship between transcripts and their encoded proteins.

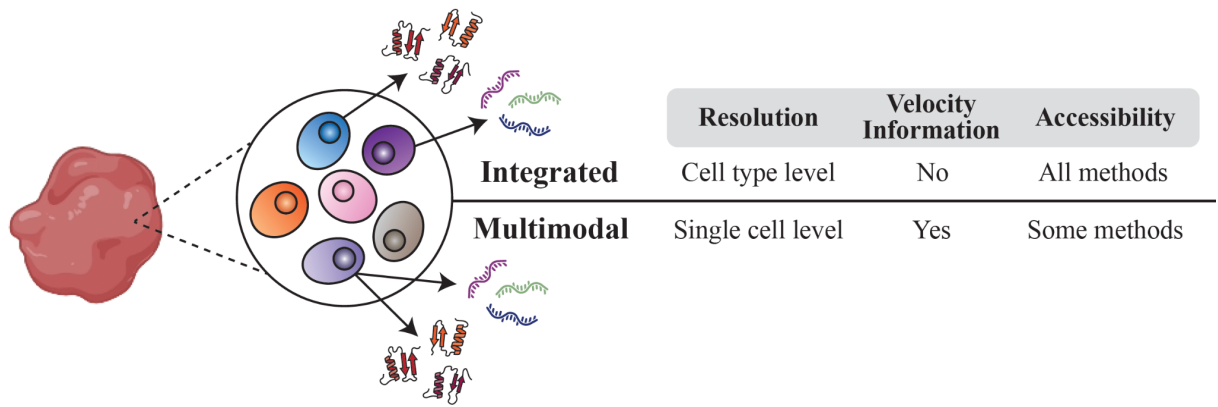

| | Resolution | Velocity Information | Accessibility |
|---|---|---|---|
| **Integrated** | Cell type level | No | All methods |
| **Multimodal** | Single cell level | Yes | Some methods |

**Figure 1 | Tradeoffs of integrated and multimodal proteogenomic measurements.** Integrated analysis refers to measuring protein or RNA in single cells independently and computationally aligning the samples, while multimodal refers to measuring different classes of biomolecules in the same cell. With integrated data, resolution of analysis may be limited to the level of cell clusters due to the challenge of aligning individual cells. Yet, it allows experimenters more flexible options for sample preparation. Fewer sample preparation options exist for multimodal analysis, but this approach obviates data integration and supports single-cell resolution modeling. It also permits modeling the time delay between transcription and protein synthesis across dynamic processes (velocity information).

## Estimating reliability

Developing rigorous models requires estimation of both random or systematic measurement errors. Random noise is more easily estimated by comparing the consistency of measurements across replicates, while systematic noise can be estimated by comparisons between methods. For example, the contribution of random noise to observed fold changes between two different cell types can be estimated from non-overlapping ensembles of cells from each cluster, Fig. 2a. Specifically, divergence of the corresponding distributions for each ensemble reflects random noise, denoted with $\Delta\varepsilon_2$ in Fig. 2b.

### Estimating random and systematic errors

Systematic errors are harder to estimate, especially when measurement methods have high precision but low accuracy. Systematic deviations can be identified by comparing the consistency between measurement methods that do not share biases, $\Delta\varepsilon_1$ in Fig. 2b. For example, high dropout rates for droplet single-cell sequencing methods may overestimate or underestimate fold changes between cell clusters depending upon how data are processed[31]. Such systematic biases may be estimated and managed by performing independent measurements, e.g., SMART-seq3 or RNA-fluorescence in situ hybridization (FISH)[32]. Similarly, systematic ratio compression of protein fold changes due to coisolation of isobarically labeled peptides may be detected and accounted for by performing quantification that is not affected by coisolation, such as plexDIA[15,33]. Additionally, depending on data acquisition strategy, multiple data points that have different interferences can be obtained from the same peptide before and after the fragmentation process required for identification[16]. More generally, MS offers multiple opportunities for detecting systematic errors. For example, proteins can be digested by different enzymes or their peptides labeled with different mass tags. Different proteases and mass tags produce different peptide ions from the same protein, and these help identify and mitigate systematic errors due to interferences. The consistency between measurements of peptides mapping to the same protein can provide largely independent data points in the context of inferences, though proteoforms have to be accounted for[34,35].

## Estimate reliabilities

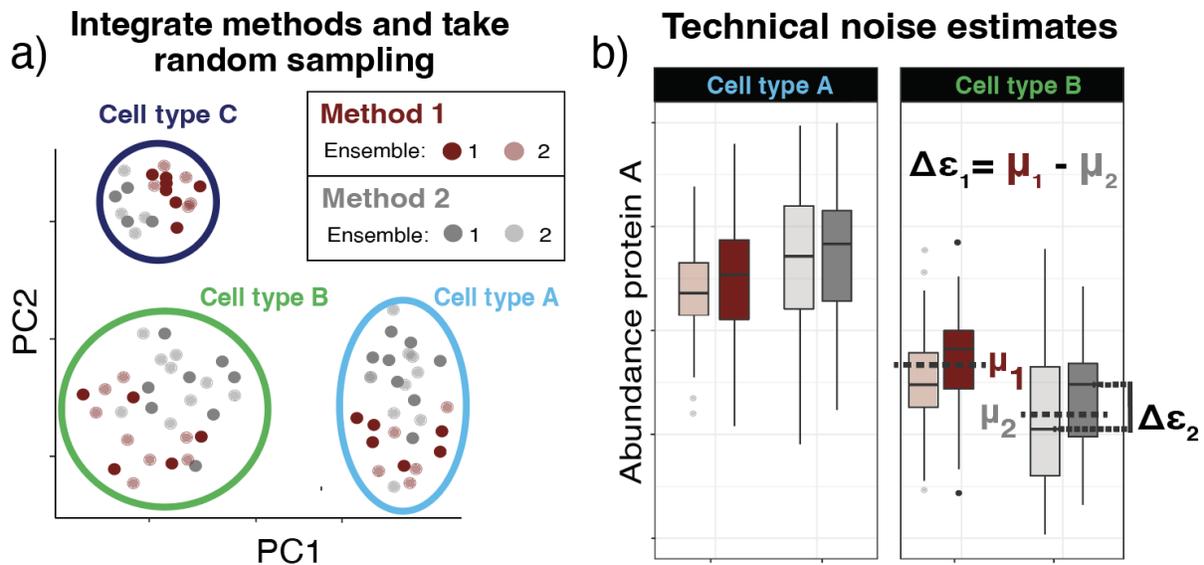

**Figure 2 | Estimating random and systematic errors.** Quantifying the degree and nature of noise in single-cell measurements is key to developing accurate proteogenomic models. This noise may be random, such as poisson noise from sampling few copy numbers of molecules, or systematic due to biases in the measurement approach. **a**, To estimate the extent and nature of noise in the measured abundances of RNA and proteins, measurements should be repeated using methods that share as few biases as possible. Once such data are collected, cell types can be aligned across different methods, and different ensembles of cells can be selected both within and across methods. **b**, Comparing protein abundances between different ensembles of cells taken from different methods(from cell type B) allow for assessing contribution of variance due to method specific bias ($\Delta\epsilon_1$), while comparing abundances between ensembles within a method allows for measuring the influence of random errors from within method variation ($\Delta\epsilon_2$).

Different computational workflows can also contribute to different sources of systematic bias[36–38]. For example, when performing analysis on integrated datasets, inaccurate alignment can result in each cluster having a different cell type representation across datasets, which can be misinterpreted as post-transcriptional regulation. Thus, models should also account for the reliability of integration, e.g., by comparing the consistency of clustering the same group of cells utilizing different subsets of gene products[39].

# Inferring regulation

Biological regulation may be inferred by models at varying levels of abstraction and mechanistic resolution, Fig. 3. Simple models may estimate systematic discrepancies in mRNA and protein abundance that exceed measurement noise (Fig. 3a). These models can reveal post-transcriptional regulation associated with phenotypic states or with functional activities of individual cells, such as the association of endocytic activity with post-translational modifications by proteolysis[40]. Such associations may lead to new hypotheses that can be probed deeper by follow up perturbation experiments. More detailed biophysical models may discriminate between different topologies of signaling networks[9] and infer associated rates of molecular interconversions (Fig. 3b). These models are more directly interpretable but require either more direct measurements or more assumptions.

## Abstract models

The simplest model connecting different layers of gene regulation can be framed as hypothesis testing; for example, testing whether the protein products of a gene are determined solely by its RNA products or not. This hypothesis testing abstracts the molecular mechanism of post-transcriptional regulation into a constant, e.g., the protein to RNA ratios (PTRs), Fig. 3a. Similarly, models can test hypotheses about RNA isoforms or modifications, such as pseudouridylation, affecting protein abundance. By explicitly modeling sources of error due to measurement, models can distinguish between instances that have similar magnitude of PTR variation but have different biological interpretation, Fig. 3a. Such abstracted models can be constrained either by integrated or multimodal data (Fig. 1) and have been applied to bulk[17,41] and single-cell proteogenomic data[20,42], however measurement reliability was not accounted for in all cases.

Factor analysis represents another abstract type of modeling that can incorporate known experimental structures, such as sparse interactions and spatial information [43–45] When applied to single-cell proteogenomics data, factor analysis can identify cell-type specific modules of transcripts covarying with their corresponding transcription factors.

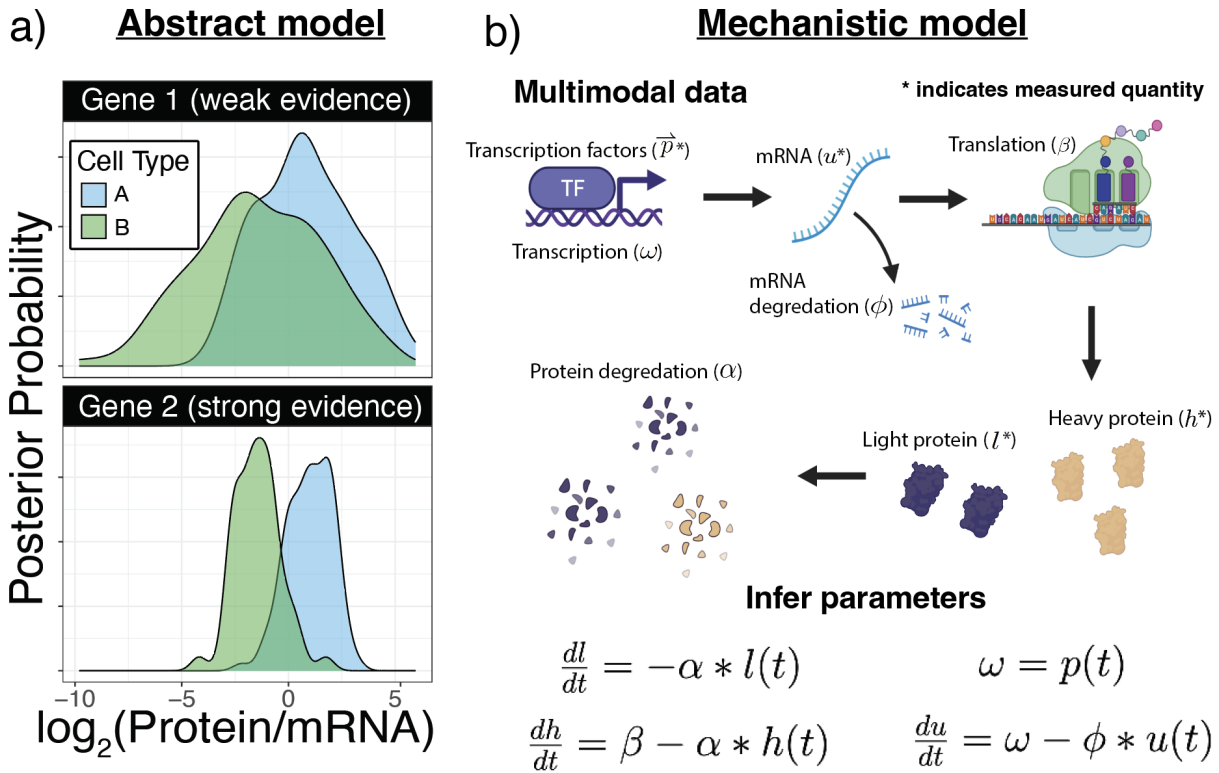

**Figure 3 | Regulatory processes can be modeled at different levels of abstraction and mechanistic resolution. a**, At a high level of abstraction, agreement between protein and mRNA levels across cell types can be explored through statistical frameworks. Such models allow for discrimination between biological and technical differences in RNA and protein levels. This discrimination requires modeling experimental and computational sources of error. **b**, To represent biological mechanisms, changes in parameters such as transcription factor abundance, protein synthesis, and degradation can be modeled in time. This allows for more interpretable and direct biophysical understanding of gene regulation. Parameters such as protein synthesis can be inferred from experimental designs that utilize metabolic labeling pulse-chase experiments to distinguish newly synthesized proteins.

## Mechanistic models

While abstract models may elucidate high level instances of post-transcriptional regulation, understanding general principles of which factors strongly contribute to post-transcriptional regulation requires more finely resolved models. To enhance the interpretability, abstract models can incorporate more realistic molecular mechanisms, Fig. 3b. For example, instead of modeling post-transcriptional regulation with a single constant, its stages (protein synthesis and degradation) can be modeled by detailed rate equations that explicitly factor in regulators, such as microRNAs, RNA binding proteins, and ubiquitin ligases. Such mechanistic models generally benefit from prior knowledge[46] and from multimodal measurements, which allow reducing assumptions and encoding temporal information. This information is implicit in proteogenomic data due to the time delay between RNA and protein dynamics[47,48], and can be made more explicit and quantitative by pulse-chase metabolic labeling RNA with 4-thiouridine or proteins with amino acids coded by stable isotopes. Such approaches enable direct measurements of protein synthesis and degradation rate, and may provide reliable enough time information to support causal inference.

Long-standing interest in mechanistic modeling of gene expression has contributed to elegant models whose applications have been circumscribed by the limited scope of protegenomic data. For example, fluorescent measurements of a few proteins allowed studying the influence of transcription factor (TF) binding on downstream protein production in single cells[49,50]. Yet, these pioneering experiments were limited to a few proteins in engineered cells. The data did not support modeling latent variables that regulate protein levels, such as TF modifications, RNA levels, and rates of protein synthesis and degradation. These factors may now be measured at increasing scales[51] or modeled directly and related to the abundance of RNA binding proteins and ubiquitin ligases. Thus, emerging technologies may provide proteogenomic data with the potential to support more comprehensive and mechanistic models while simultaneously extending the analysis to thousands of gene products across diverse samples, including from human patients.

Proteogenomic data parameterized in absolute units, such as copies of protein molecules per cell or molar concentrations, are essential for answering many questions. For example, understanding the specificity and kinetics with which limited copies of a transcription factor bind across the genome[52]. Such parameterization can be achieved when proteins are quantified by MS relative to spiked in standard of proteins or peptides with known absolute abundance. With such data, the models from Fig. 3b can reveal new regulatory principles, such as how the mode and dynamics of regulation depend on protein copy numbers or turnover rates.

Proteogenomic models defined in terms of direct molecular interactions quantified by biophysical parameters are likely to be more generalizable. Unlike indirect associations that vary from dataset to dataset, molecular interactions provide more invariant representations of biological systems that provide (i) more specific testable hypotheses and (ii) more robust predictive models. The estimation of model parameters should explicitly account for measurement noise to avoid biases leading to overestimation or underestimation of parameters[53]/

## Examples of proteogenomic analysis

Single-cell proteogenomic has the potential to elucidate post-transcriptional across a broad range of biological problems. Here suggest a few examples that illustrate different domains of biomedical applications.

## Inferring protein interactions from buffered DNA copy number variation

Deletions or duplications of DNA are detected in healthy human cells[54] and common in cancer cells due to their genome instability[55,56]. Such DNA copy number variation (CNV) is reflected in transcript abundance variation but substantially buffered at the protein level, especially for proteins forming complexes[57]. Thus, the extent to which a protein is buffered against increased or decreased copy gene number in a single cell can suggest whether that protein is a member of a complex. This information, combined with shared protein covariation between complex members[9], can empower inference of protein complexes in the cellular subpopulations of human tissues. Since CNV can be inferred from RNA abundance measurments[56], this analysis can be extended to single-cell proteotranscriptomic datasets, including from patients.

While such inference from CNV can be applied to many systems and diseases, it is likely to be particularly informative in cancer samples where genome instability results in more CNV variation across the genome and single cancer cells. This extensive CNV variation and its attenuation across the single cells can indicate changes in protein complexes between normal and cancer cells, including complex rewiring that alter protein functions and signal transduction[58]. This information may reveal activated and inhibited signaling pathways that can be therapeutically targeted.

## Inferring transcriptional regulation to improve directed cell differentiation

A clear application of single-cell proteogenomic data is the inference of transcription factors (TF) regulating transcription. Such inference has been of intense interest since the advent of DNA microarrays in 2000s[59] and received renewed interest with the widespread adoption of single-cell RNA sequencing[60]. Yet inferences from different approaches using the same single-cell RNA data provide divergent results[61]. These challenges are due in part to the fact that the abundance of RNAs coding for TFs are incomplete surrogates for the active forms of the TFs. Modeling TFs as latent variables can partially mitigate this challenge[43,62], but direct measurements of post-translationally modified TFs in the nucleus can enable more accurate inferences. This is particularly the case if all relevant TF are measured and the reliability as measurements estimated (Fig. 2).

This inference requires global proteogenomic measurements (since mammalian genes are regulated by many TFs) and can benefit many applications, such as rational optimization of directed stem cell differentiation. Indeed knowledge of TF-target genes informed the first induction of pluripotency by Yamanaka and colleagues[63], and such knowledge remains useful and limiting for improving directed stem cell differentiation. This limitation may be relieved by regulatory inferences from single-cell proteogenomic data, which will facilitate rational engineering of targeted differentiation for regenerative therapies.

# Conclusion and outlook

Maturing MS technologies will add a new modality to single-cell multimodal measurements, and thus contribute to the emergence of single-cell proteogenomics. The quantitative accuracy and reliability of these measurements will empower biological models proving generalizable and predictive representations of biological systems. These models will help quantify and interpret differences between single-cell modalities.

**Acknowledgements:** We thank Aleksandra Petelski and other members of the Slavov laboratory for discussions and comments. This work was supported by an Allen Distinguished Investigator award through The Paul G. Allen Frontiers Group, and an R01 award from the National Institute of General Medical Sciences from the National Institutes of Health to N.S. (R01GM144967).